\documentclass[12pt]{article}

\usepackage{a4wide}
\usepackage{authblk}
\usepackage{graphicx}
\usepackage{epstopdf, epsfig}	
\usepackage{amsmath}
\usepackage{amssymb}
\usepackage[round,sort&compress,comma,authoryear]{natbib}
\usepackage{float}
\usepackage{hyperref}
\usepackage{lscape}
\usepackage{xcolor}
\usepackage{newtxtext}
\usepackage{newtxmath}
\usepackage{natbib}
\usepackage{hyperref}
\usepackage{fdsymbol}
\hypersetup{
    colorlinks = true,
    urlcolor   = blue,
    citecolor  = black,
}
\definecolor{front_jet}{RGB}{128, 229, 255}
\definecolor{second_row_jet}{RGB}{0, 170, 212}
\definecolor{third_row_jet}{RGB}{0, 136, 170}
\definecolor{middle_jet}{RGB}{0, 68, 85}
\definecolor{bulk}{RGB}{0, 85, 212}

\newcommand{\jon}{\textcolor{black}} 
\newcommand{\naren}{\textcolor{black}} 

\title{Air-water cavity in multi-plunging jet impacts}

\author[1]{Narendra Dev}
\author[1]{H\`{e}l\'{e}ne Scolan}
\author[1, $\dagger$]{{J. John Soundar Jerome}}
\author[1]{Jean-Philippe Matas} 
\affil[1]{Universit\'{e} Lyon 1, LMFA, UMR5509, CNRS, \'{E}cole Centrale de Lyon, INSA Lyon, 69622 Villeurbanne France}
\affil[$\dagger$]{email : john-soundar@univ-lyon1.fr}

\date{}
\begin{document}

\maketitle

\begin{abstract}
{We report the formation of an original dome-shaped air-water cavity at the impact site of closely-packed plunging water jets. The injector assembly consists of concentric rings of circular jets, similar to a shower-head configuration. We infer from high-speed imaging, LASER-induced fluorescence, and optical phase detection probes that the cavity consists of multiple stems and sheets of air that stretch, retract and pinch off to produce bubbles. We also illustrate various bubble production mechanisms in the cavity, along with coalescence and bubble breakup scenarios in the subsequent bubble cloud. Thereby, we describe the wide range of bubble sizes generated by such jets, from a few micrometers up to about $5$~mm. Measurements of the cavity size are carried out for varying impact velocity, number of jets, and jet spacing. We propose that this distinctive air-water cavity is generated by liquid entrainment when successive rings of plunging jets progressively mix with the pool water.}
\end{abstract}

\section{Introduction}
\label{sec:headings}

Air entrainment into liquids by plunging solids \citep{aristoff2009} and liquids \citep{bin1993, kiger2012air} is a very common phenomenon. It is customary to employ freely-falling droplets and jets impinging on a quiescent pool to investigate air-liquid cavity at the impact site in search of air entrainment mechanisms. For a smooth viscous jet \citep{lin1966gas}, beyond a visco-capillary critical velocity \citep{eggers2001air, lorenceau2004air}, a long, thin and trumpet-shaped sheath of air is formed around the jet as it expands into the pool. This air-film is known to wobble and break down into bubbles.  However, when viscous effects are not predominant, the entrainment process is strongly influenced by perturbations in both the bulk flow and the jet morphology. \citet{mckeogh1981air} observed the so-called \textit{intermittent vortex regime} in a rippled jet. A deep asymmetric air cavity persists, somewhat erratically, around the expanding jet, and fragments into bubbles at its tip. Many subsequent studies \citep[see Figures 5--8, and references therein]{kiger2012air} also observed air cavities in the form of \textit{stems} and \textit{fingers} around the plunging jet. {Such} structures are expected to be generated on the free surface during the impact of either a weakly-perturbed jet due to the action of strong azimuthal rotation, similar to a vortex-ring.  {Similarly, deep and narrow cavities entrap bubbles while collapsing under gravity, if a \textit{bump} in the jet profile is introduced via a local flow-rate variation \citep{Ohl2000, zhu2000mechanism}.} In fact, vigorous air entrainment occurs when the jet morphology is highly perturbed as first observed by \citet{lin1966gas}, and many others afterwards \citep{van1976, ERVINE1980, sene1988air, bonetto1994analysis, Davoust2002, dev2024, Redor2025}. {This is also because jets often fragment during free-fall and disintegrate into droplets \citep{lin1998drop}. Recently, \citet{bouwhuis2016impact, speirs2018water, lee2021elongated} investigated the air cavity dynamics during the impact of a series of drops. They observe that a deep and narrow, elongated cavity is generated initially. Then the cavity collapses into a funnel of a few millimeters in depth, depending on the droplet size and impact frequency \citep{lee2021elongated}.}


\naren{In addition to single plunging jets and droplet trains, natural flows such as waterfalls often involve large jets fragmenting into multiple continuous and discontinuous streams and droplets. Configurations of this type arise in hydraulic structures such as stepped spillways and cascades \citep{chanson2002hydraulics}, as well as in aeration and mixing devices used in wastewater treatment \citep{ohkawa1986flow}. 
An idealized ensemble of continuous, parallel jets could therefore be employed as a controlled and reproducible configuration to investigate free-surface deformation and the associated liquid and air entrainment mechanisms in multi-jet impingement. Although such tightly packed continuous jets do not occur directly in natural flows, establishing an understanding under idealized conditions is a necessary first step, as these mechanisms would be exceedingly difficult to isolate and interpret in realistic flows involving multiple impact zones in severely fragmented jet. Within this context, previous studies on identical multi-plunging jets have primarily focused on oxygen transfer rates and efficiency.}  For example, \citet{van1981water} examined two inclined jets at a constant velocity and reported that the volumetric flow rate of air entrained into the water was reduced by up to $10$\% compared to a single jet. Subsequently, \citet{tojo1982} conducted experiments on twin plunging jets and reported a pronounced dependence of oxygen mass transfer and mixing time on the jet angle, achieving an optimum at $\pi/3$ under fixed spacing conditions.
Later, \citet{ide2003} studied $N = 3$ jets containing fine dissolved bubbles and demonstrated an increase in the maximum penetration depth of the bubble cloud $H$ and gas hold-up with the number of jets ($N$) at constant volumetric flow rate per jet. \citet{deswal2007} performed experiments with up to $N = 16$ jets of equivalent total cross-sectional area to a single jet. At fixed jet velocity, the volumetric oxygen transfer coefficient increased with $N$ due to the larger interfacial area exposed to the atmosphere, with efficiency gains up to $1.6$ times relative to single jet configurations. \citet{mishra2018} reported that for a given total flow rate, $H$ decreased as $N$ increased. They attributed this effect to the reduced momentum of individual jets and greater frictional losses associated with multiple jet interactions. More recently, the present authors \citep{dev2025} considered the case of hexagonally-packed array of circular plunging jets, with up to 61 parallel jets. Here, the increase in air entrainment rate as a function of the number of jets was proportional to the perimeter of the composite  multi-jet. Studies on such multi-jet configurations and the cavity created at impact could  provide crucial insights into the mechanisms of air entrainment and oxygen transfer. They are potentially relevant to highly fragmented large-scale jets as well. Unlike the single-jet case, the biphasic cavity just below impact for an ensemble of jets has never been considered before. This is precisely the goal of the present work. We study experimentally the impact region for varying multi-jet configurations, analyse its biphasic structure and size, and obtain critical insights for bubble production, coalescence and break-up.

\section{Cavity at the impact of a multi-plunging jet}
\label{sec:cavity_exp}
\begin{figure}[!h]
  \centering
  \epsfig{file=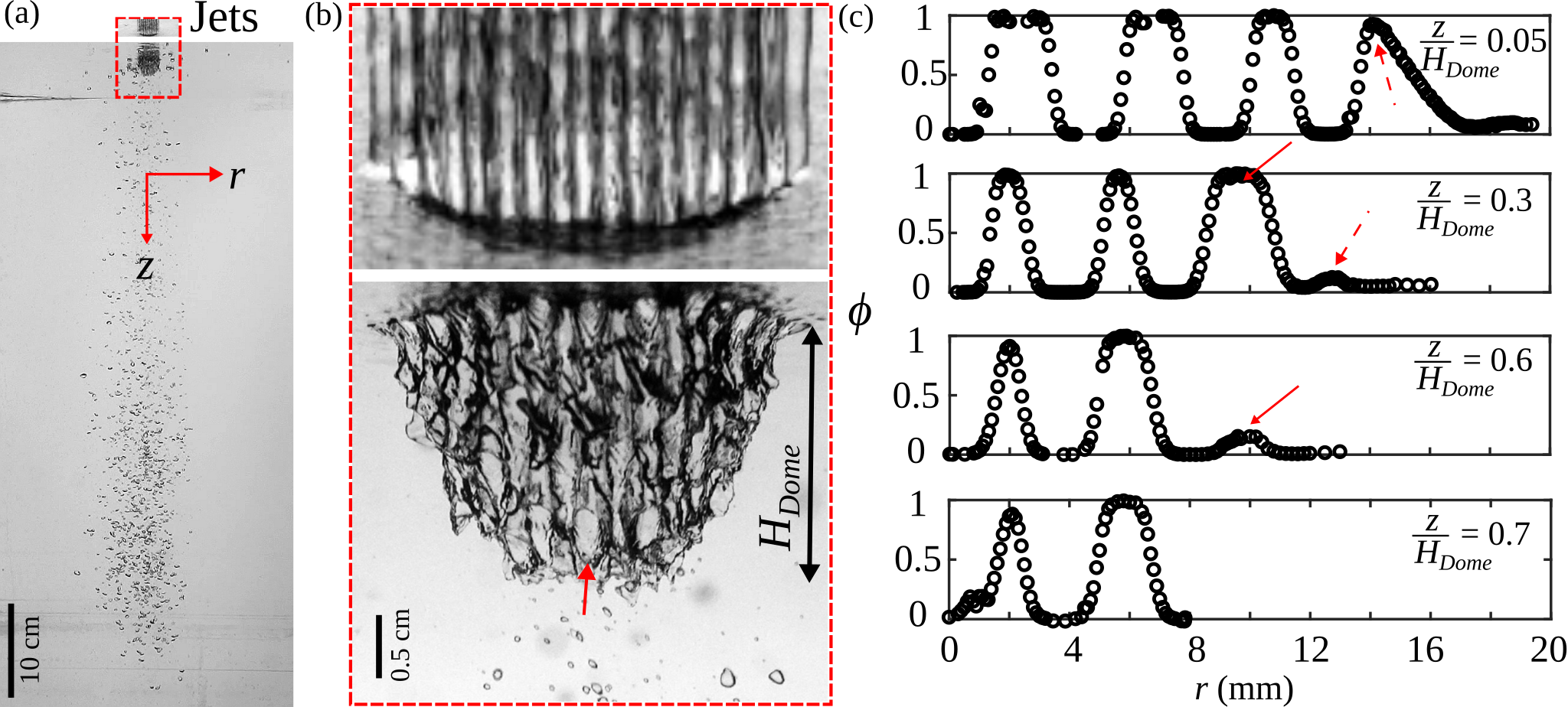,width=1\textwidth,keepaspectratio=true}

  \caption{{(a) {Concentric hexagonal rings of $61$ circular jets ($D_n = 2.7$~mm)} plunge in a water pool at $V_i = 1.4$~m/s to form a dilute cloud. Height of fall $Z_f = 2$~cm. (b)  Plunging jets just above the impact (top-row) and cavity just below the free surface (bottom-row). (c) Radial profiles of the void fraction $\phi$ at four different depths, $z$ taken along the center-plane of the cavity. They show alternating water tubes ($\phi = 0$), and air stems ($\phi = 1$) inside the cavity.}} 
  \label{fig:dome61}
\end{figure}

Figure~\ref{fig:dome61}(a) displays a \textit{cloud} of bubbles generated by plunging an ensemble of $61$ jets into water contained in a glass tank ($1.2$~m deep $\times$ $0.6$~m $\times$ $0.6$~m). \naren{ Jets are arranged in concentric hexagonal rings such that the distance between the closest neighbours is kept constant at $X = 1.5 D_n$, where $D_n = 2.7$~mm is the diameter of identical holes in the injector plate (see Figure \ref{fig:CAD}c).}  \naren{The fall height is $Z_f = 2$~cm, and the impact velocity $V_i$ of each jet is deduced from the nozzle exit velocity $V_n$, obtained from the flow rate measured using the flow-meter, using $V_i = (V_n^2 + 2 g Z_f)^{1/2}$}.  The  water level is maintained by slightly tilting the reservoir on one side so that overflowing water is drained to a secondary reservoir bucket, as in \cite{dev2025}.
Such closely-packed jets remain roughly parallel without merging along their fall for these conditions, as observed in Figure \ref{fig:dome61}(b). After impact, the jets entrain air to about {$64$~cm} deep into the pool. A closer look at the impact site is provided in {Figure \ref{fig:dome61}(b) (bottom)}. Here, the ambient air is observed to be entrained well below the water surface  so as to form an \textit{inverted} dome. We invite the reader to watch the {Supplemental video I} showing time resolved images acquired with a \textit{Photron SA4} camera. \naren{All the movies in this study are captured at 10,000 and 20,000 FPS for jets and cavity, respectively.} Multiple air cavities, in the form of thin sheets, stems or filaments, are visible around arrays of water tubes. {The latter} become longer near the center of the jet ensemble as the jets penetrate deeper into the pool and mix with the bulk water. Oval openings can be seen near the bottom of the structure, as indicated by the small arrow in {Figure \ref{fig:dome61}(b) (bottom)}.

The air and water content in the cavity can be further explored with the help of phase-detection optical probes (A2 Photonic Sensors). Figure \ref{fig:dome61}(c) displays radial measurements of the time-averaged air-to-water ratio, hereafter referred to as void fraction $\phi$, at different axial locations with a spatial resolution of $0.1$~mm in the cavity formed by $N$ = 61 jets. 
All data present strong spatial variation of the void fraction between zero (water) and unity (air). Intermediate values of $\phi$ represent a mixture of the two phases.   The measurements taken just at the impact, i.e., $z/H_{Dome} = 0.05$, reveal five jets ($\phi$ = 0)  and only four air signals ($\phi$ = 1), since the outermost jet mixes with the bulk liquid, preventing the formation of a localised air region beyond this point. As $z/H_{Dome}$ increases, the outermost air signal diminishes, as indicated by the low $\phi$ values (marked by red dashed and solid arrows). As we move towards the nose of the inverted dome-shaped cavity, the number of peaks and troughs decrease. This suggests that fewer air pockets surrounding water jets are present inside the cavity at larger depths, consistent with {Figure \ref{fig:dome61}(b)}. Troughs ($\phi = 0$) corresponding to signals from the liquid phase appear slightly wider than peaks ($\phi = 1$). \naren{Probe signals also present a very small peak in void fraction due to air entrainment at the outer-most edge of the multi-jet. This region is similar to the funnel-like cavity in the case of a single jet when the free-surface dips into the pool at the jet periphery.} \jon{The small radial shift of jet positions between stations $z=0.05~H_{Dome}$ and $z = 0.3~H_{Dome}$ is due to a slight misalignment in the plane of measurement.}

\begin{figure}[!h]
  \centering
  \epsfig{file=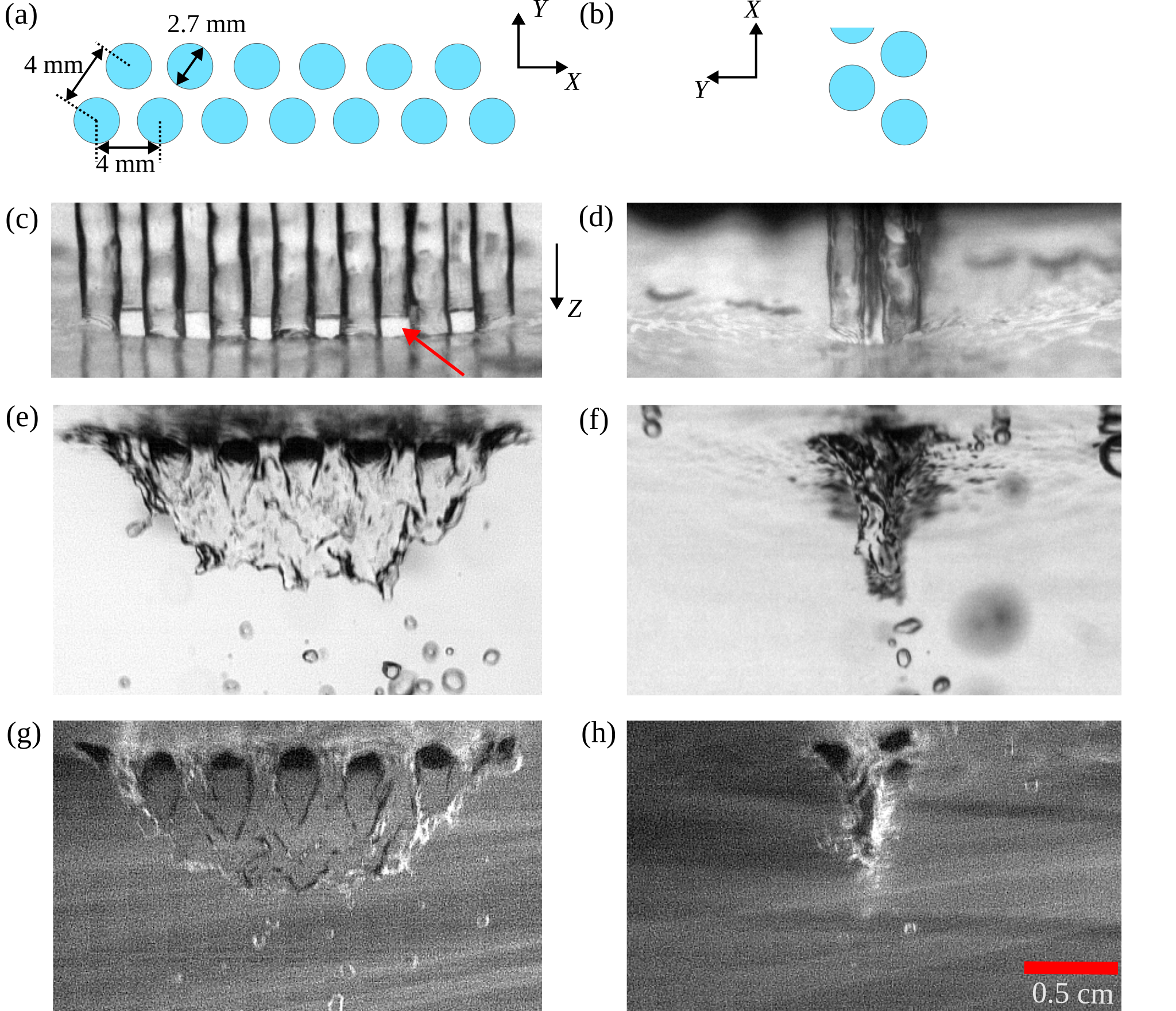,width=0.85\textwidth,keepaspectratio=true}
\caption{ \naren{Top view of the linear arrangement of $N$ = 13 jets with $X/D_n$ = 1.5 as seen from (a) front view and (b) side view. Instantaneous images of liquid jets impacting the pool surface at $V_i$ = 1.5 m/s captured from (c) front and (d) side views. Back-lit images of the air cavity beneath the water surface are seen from (e) front and (f) side views. LIF images of the air cavity beneath the water surface are seen from (g) front and (h) side views. The readers are advised to watch the Supplemental video II. }} 
\label{fig:array_jet}
\end{figure}

\naren{To have a better understanding of the outer structure of the dome, we further analysed the simplified case of multiple plunging jets arranged in two rows, with seven jets in the first and six in the second row, as shown in Figure \ref{fig:array_jet}(a) (front view) and (b) (side view). Figure \ref{fig:array_jet}(c) illustrates the front view of liquid jets plunging into the pool, while Figure \ref{fig:array_jet}(d) provides a side view of the jets. Note that front and side views are not taken at the same instance. Figure~\ref{fig:array_jet}(c) reveals an illuminated region between the jets due to refraction of light at the air-liquid interface, as delineated by the red arrow. This optical signature indicates the presence of an air cavity formed between neighbouring jets, which is shown in the front and side views in Figs.~\ref{fig:array_jet}(e) and (f), respectively. In the front view, the cavity has an overall U-shaped form, whereas in the side view it appears as a slender, curved V-shape. Figure~\ref{fig:array_jet}(e) further shows alternating dark, rounded regions at the meniscus corresponding to the jet locations, with pointed oval structures visible immediately below each jet. Overall, the cavity exhibits a wrinkled morphology due to capillary effects and is composed of multiple air-filled nodules and finger-like structures along its periphery. To further understand the oval structures of the cavity, a high-speed LASER-induced Fluorescence (LIF) imaging is performed with a  $450$~nm wavelength LASER (\textit{Z-LASER Z1000-ZQ1}). A high-pass filter having a cut-off around $510$~nm  (Lee filter \textit{010-Medium Yellow}) is placed in front of the camera, in order to only detect fluoresced light. The fluorescein dye has an absorption peak of $490$~nm and an emission peak of $510$~nm. Figure \ref{fig:array_jet}(g) presents an instantaneous LIF image of the cavity wherein water is fluorescein dyed. Interestingly, although the air itself contains no dye and cannot fluoresce, the contours of the air fingers and bubbles are distinctly visible, due to the reflection and \textit{scattering} of fluorescent light at the concave air-water interfaces, which enhances their visibility. The oval structures are now more clearly resolved which  delineate the air?liquid interfaces. These features actually represent the projected images of the parts of the jets in the cavity, viewed through the surrounding bulk medium. This interpretation is consistent with the alternating air?water signal observed in the void-fraction measurements shown in Figure \ref{fig:dome61}. These projections arise from the interaction of the cylindrical jets with the cavity.}

Turning our attention back to the hexagonal jet arrangement, Fig.~\ref{fig:CAD}(a) presents an instantaneous LIF image of the biphasic cavity with fluorescein-dyed water. Similar oval structures to those observed in Fig.~\ref{fig:array_jet}(g) are again visible. Some of these structures are located close to the free surface and correspond to the outermost jets interacting with the surrounding bulk liquid. Further below, more elongated oval structures are observed, which we interpret as arising from different rows within the hexagonal jet arrangement. We advise the reader to look at {Supplemental video III} showing the associated time resolved video.
\begin{figure}[!h]
  \centering
  \epsfig{file=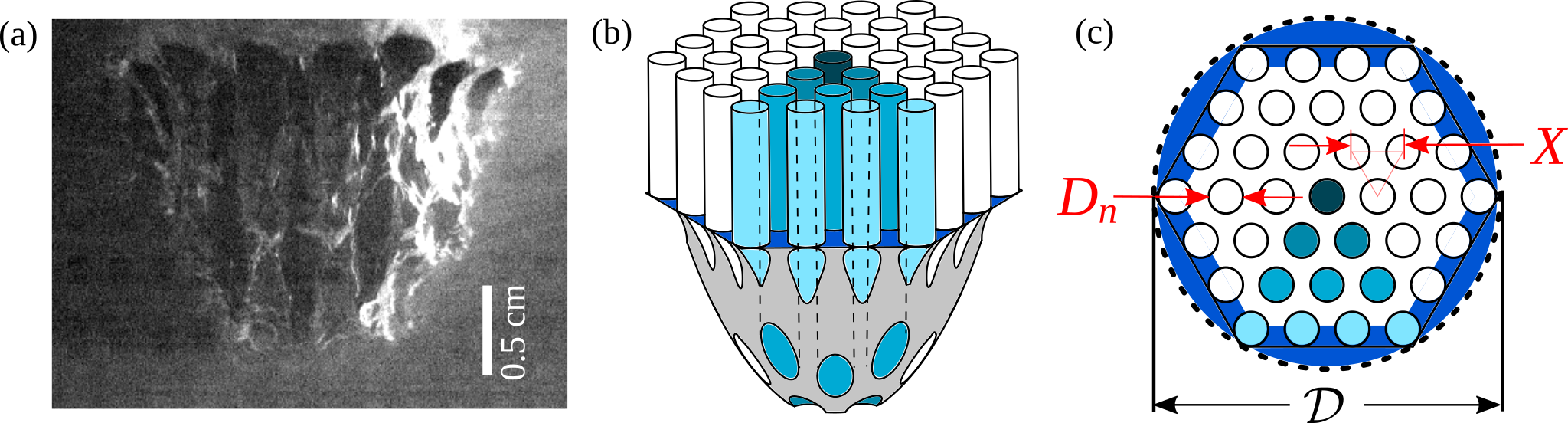,width=1\textwidth,keepaspectratio=true}
  \caption{(a) Instantaneous LIF image of the dome formed just below the surface by $N = 37$ jets at $V_i$ = 1.5~m/s and $Z_f=2$~cm. (b) A 3D schematic view of jets and the dome corresponding to the arrangement of the colour-coded circular jets from the injector plate sketched in (c) with $N$ holes of diameter $D_n$, arranged in a hexagonal pattern with center-to-center spacing $X$. Here, $\mathcal{D}$ denotes the diameter of the circumcircle. \naren{Note that the void-fraction measurement shown in Fig.~\ref{fig:dome61} is taken along the central plane marked by dashed line in (c).}}
\label{fig:CAD}
\end{figure}
Leveraging this understanding, Figure \ref{fig:CAD}(b) displays a 3D model of the cavity where the colour codes of different jets correspond to the top view of the cavity depicted in Figure \ref{fig:CAD}(c). Note that the injector plate consists of a hexagonally-packed circular holes of diameter $D_n$, see Figure \ref{fig:CAD}(c). Each holes are separated from their closest neighbour by a distance  $X$. Here, the bulk water is represented in dark blue \colorbox{bulk}{\makebox[0.1cm]{\rule{0pt}{0.1cm}}} while jets in different rows are distinguished by varying shades of steel blue. Comparing Figures \ref{fig:CAD}(a), (b), (c) the four jets in first row marked by \colorbox{front_jet}{\makebox[0.1cm]{\rule{0pt}{0.1cm}}} form the base of the dome. Similarly, the three second row jets identified by \colorbox{second_row_jet}{\makebox[0.1cm]{\rule{0pt}{0.1cm}}} form the middle elongated oval structures, again indicating that these three jets are observed through the air cavity. Both third row jets marked by \colorbox{third_row_jet} {\makebox[0.1cm]{\rule{0pt}{0.1cm}}} form the two curved surfaces at the bottom. The central jet, marked by \colorbox{middle_jet}{\makebox[0.1cm]{\rule{0pt}{0.1cm}}}, is not visible in the images.

\section{Bubble production, break-up and coalescence}
The presence of the dome strongly influences bubble production in the case of multi-jets. To illustrate this, the dome formed by $N = 37$ jets at a low impact velocity of $V_i = 1.9\ \text{m/s}$ is considered in Figure \ref{fig:bubble_production}(a). Image sequences are then provided at three different bubble production sites, indicated by boxes. Figure \ref{fig:bubble_production}(b) highlights the growth of a small air cavity at the dome periphery, indicated by the red box in Figure \ref{fig:bubble_production}(a). This small cavity develops between the outermost jets and the surrounding bulk water. \jon{This leads to the formation of a small peak at $r = 18$~mm.}The image sequence captures the elongation of the cavity from a nodule into a finger, which subsequently pinches off bubbles (circles at $t = 4.7$ and $6.35\ \text{ms}$) due to necking (arrows at $t = 4$ and $t = 6.2\ \text{ms}$) and retraction of the finger. The cavity forms when the jet surface becomes corrugated, generating transient \textit{craters} upon impact that evolve into fingers, or stems. {This mechanism is predominantly observed at higher velocities \naren{due to turbulence}, but it may also occur at lower velocities \citep{Redor2025}. As mentioned in the introduction, this mechanism is the dominant one in the case of low viscosity, high-speed single jets \citep{kiger2012air}.
\begin{figure}[!h]
  \centering
  \epsfig{file=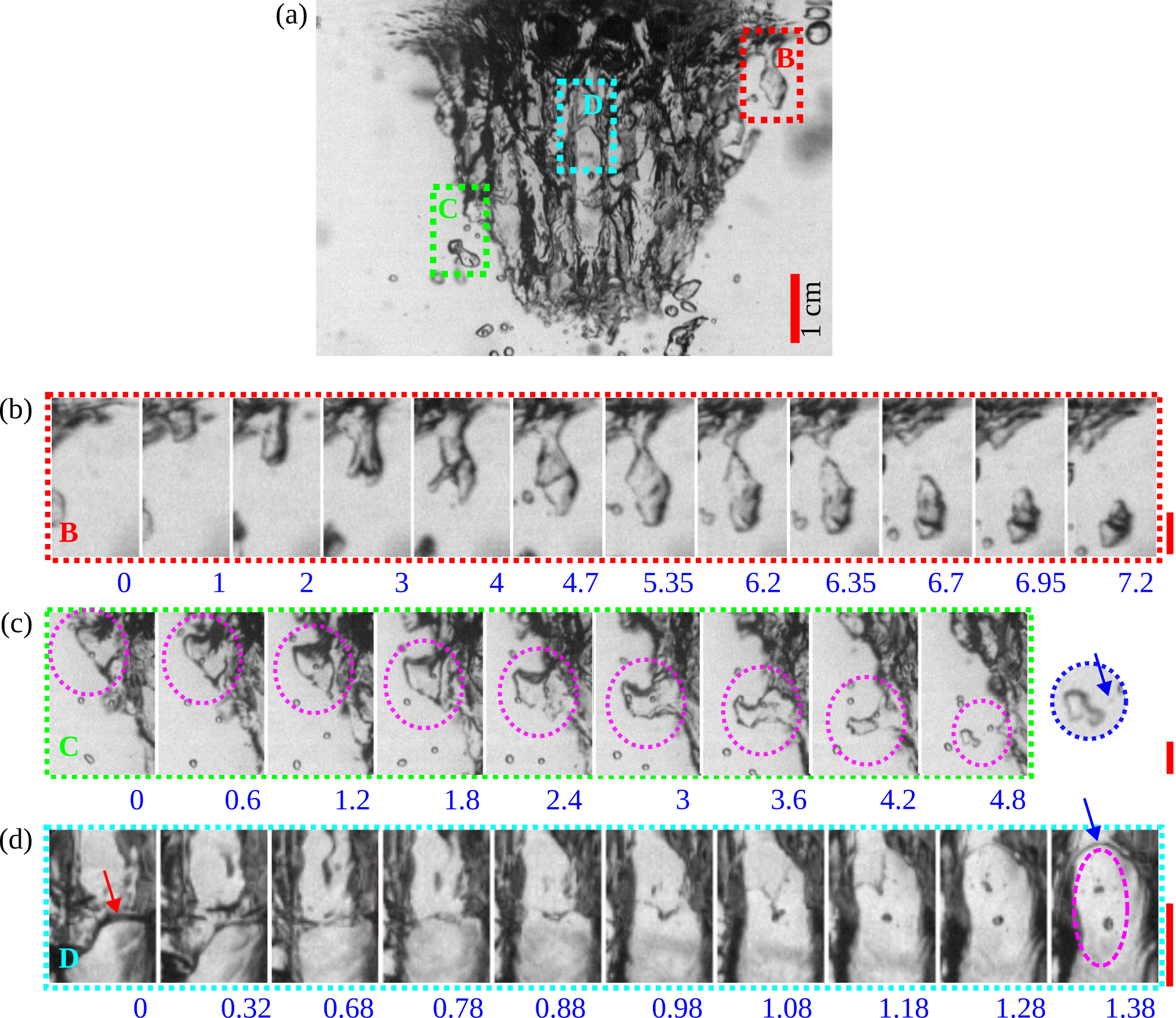,width=1\textwidth,keepaspectratio=true}

  \caption{(a) Photograph of the dome formed by $N$ = 37 jets. 
  {Image sequences showing (b) bubble formation at the periphery, close to the outermost jets, (c) evolution of an air nodule into a finger undergoing necking which results in bubble pinch-off, and (d) retraction of a narrow air sheet leading to its breakup and formation of tiny bubbles. The two arrows show initial and final interfaces, respectively. The scale bar for the image sequences is $2$~mm and the time stamps are in milliseconds.}}
  \label{fig:bubble_production}
\end{figure}
Figure \ref{fig:bubble_production}(c), corresponding to the green box location, illustrates how smaller bubbles can be formed in the present case: in this image sequence, the surface irregularity is located at a deeper location. It grows from nodule to finger as in (b) but when it undergoes necking and retraction ($t$ = 3 to 4.2 ms)  small bubbles can be pinched off the finger tip ($t$ = 4.8 ms). 
In some cases a micron size bubble is also formed as shown by the arrow in the magnified view of frame at $t$ = 4.8 ms.  Such air fingers around the dome are seen for every jet velocity but they only shed bubbles after a certain threshold. Below this velocity the fingers mostly only stretch and retract, and occasionally shed micron-sized bubbles. Beyond the threshold, the air-fingers {regularly} pinch off small bubbles. The interested reader can see {Supplemental video IV} to observe these effects of impact velocity on bubble production.

The injector geometry is such that the air cavity between two hexagonal rings of liquid jets  undergoes stretching, thereby forming an air sheet. The previously observed oval ends correspond to the rim of this air sheet, which is stationary at the depth where the jets mix with the liquid in the pool. This is illustrated by the cyan box in Figure~\ref{fig:bubble_production}(a). Figure~\ref{fig:bubble_production}(d) illustrates the temporal evolution of such a rim within this box, highlighted by the arrow at \(t = 0~\text{ms}\). Due to surface tension, the rim shrinks primarily as a result of drainage along its length, while its position remains initially unchanged ($t < 0.9$~ms). This process leaves behind a thin rim which retracts, and subsequently, breaks within one-tenth of a millisecond, to generate a micro-bubbles smaller than $200$~$\mu$m in diameter ($t = 1.18$~ms). As retraction continues, the rim undergoes successive breakup into even smaller bubbles, as observed between $t = 1.18$~ms and $t = 1.38$~ms {(see Supplemental video V)}. 
The formation and breakup of stretched air films into micro-bubbles have previously been reported in studies of droplet impact on deep pool surfaces \citep{Esmailizadeh1986, Thoroddsen2012}, where the air film becomes entrapped and stretched between two curved interfaces. More recently, \citet{Chan2019} demonstrated through numerical simulations that similar micron-sized bubbles are generated in breaking waves due to the entrapment and stretching of air films between interfaces. In contrast, the present study identifies a distinct mechanism, wherein the air film is opened from one end rather than being confined between liquid interfaces. The resulting micron-sized bubbles do not experience appreciable buoyancy and are instead carried downward by the incoming jets. In the broader context of natural water bodies, micro-bubbles are of considerable significance, as they contribute toward sustaining aquatic life \citep{khuntia2012microbubble}. 

Whereas the previously mentioned pathways for cavity break-up and the subsequent bubble formation are crucial for air entrainment rate, the bubble size distribution in plunging jets will also be impacted by break-up and/or coalescence. 
In fact, \citet{dev2025} combined phase-detection and Doppler effect in optical probe signals to elucidate that the average chord length of bubbles increases from $500$~$\mu$m up to $1.5$~mm as one moves deeper along the bubble cloud. At the foot of the cloud, bigger bubbles then escape from the expanding bubble-laden jet and rise towards the free surface. Figure \ref{fig:coal_break}(a) is an instantaneous image of the cavity formed by $N = 37$ jets at $V_i = 2.4 \, \text{m/s}$: for this velocity, larger than in the previous figures, a large number of small bubbles are produced all around the surface of the dome, and are clearly visible in the lower central region. Much larger ellipsoidal rising bubbles are visible around the dome. {This size difference arises from multiple coalescence events occurring within the belly of the bubble cloud as it slows down.}
\begin{figure}[!h]
\centering
  \epsfig{file=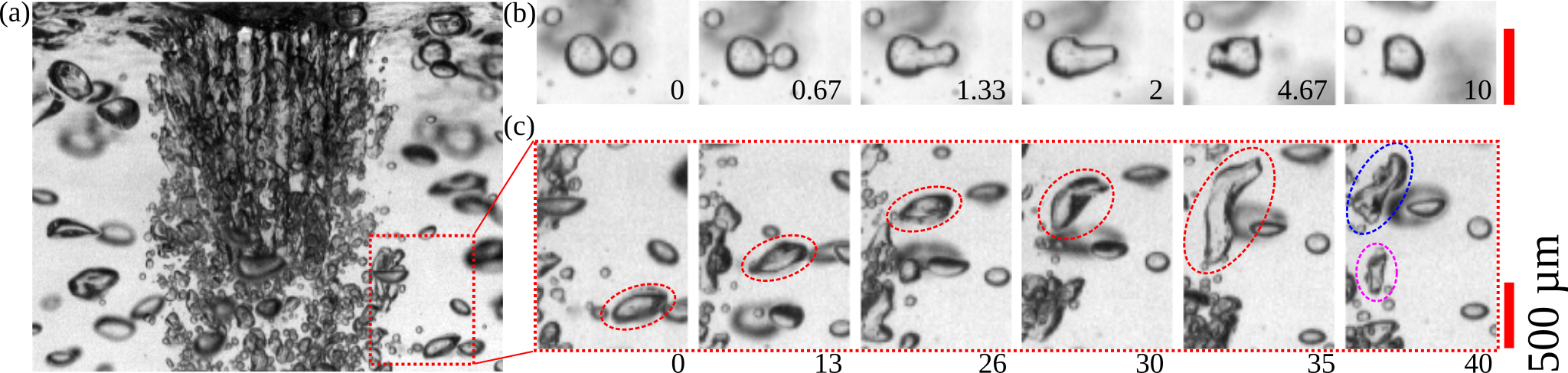,width=1\textwidth,keepaspectratio=true}

  \caption{(a) Instantaneous image of the cavity formed by $N=37$~jets at $V_i = 2.4 \, \text{m/s}$ (b) Photographs of a typical bubble coalescence sequence in the bubble cloud. (c) Shear breakup of a rising bubble. The scale bar is $5$~mm.}
\label{fig:coal_break}
\end{figure}
An example of a binary collision leading to coalescence is illustrated in the image sequence in Figure \ref{fig:coal_break}(b). While such coalescence events are expected to be frequent within the bubble cloud, rising bubbles might also interact with the shear layer and get re-entrained by the jet. Figure \ref{fig:coal_break}(c) presents an image sequence of the region of shear layer just below the dome, indicated by the box in Fig. \ref{fig:coal_break}(a). Here, a rising bubble is progressively pushed towards the shear layer ($t = 0$--26 ms). Although buoyancy prevents the bubble from being entrained downward by the liquid in the jet, the interaction induces bubble elongation ($t$ = 30--35 ms). This elongation results in necking and eventual breakup into two daughter bubbles, as marked by circles at $t$ = 40 ms. The smaller daughter bubble is transported into the shear layer, whereas the larger one undergoes further elongation and breakup cycles, ultimately leading to the complete transport of the initial bubble back towards the bottom (not shown here, see {Supplemental Video VI}). 

\section{Cavity shape and height}
As already depicted in Figure \ref{fig:CAD}(c), the injector plate consists of multiple orifices, each of diameter $D_n =2.7$~mm, arranged in a concentric hexagonal pattern with jet spacing $X$. For the conditions of interest here, the spacing $X$ varies from $1.25D_n$ to $2D_n$ and the impact velocity, based on free-fall, ranges between $1$ and $4$~m/s. The corresponding Reynolds numbers based on jet diameter $D_n$ is in the range of $2 \times 10^3$ --$10^4$. The number of active plunging jets is adjusted by selectively occluding the injector orifices with adhesive tape. \naren{Thereby, the effective diameter $\mathcal{D}$ of the multi-jet varies between $5$~mm and $5$~cm, as the number of jets $N$ varies from $3$ to $91$}. A heavy-duty centrifugal pump (2KVC AD 45/80M) from DAB Pumps and float flow meters (Georg Fischer SK 11 and SK 21) are used to maintain a constant flow rate, as required for different experimental conditions.

Figure \ref{fig:N_Dome} illustrates the evolution of the cavity with increasing number of jets ($N$) for a fixed spacing of $X/D_n = 1.5$, at various impact speeds, $V_i$. Note that, in the case of a single jet \citep{kiger2012air}, the cavity is a symmetric cusp for high-viscosity jets and an asymmetric hollow, or biphasic rim around an expanding jet for low-viscosity conditions. {As described in previous sections, we observe once again in Figure \ref{fig:N_Dome} that the cavity of a multi-plunging jet develops an \textit{inverted} dome-like shape which contains air stems, sheets, and bubbles.}
\begin{figure}[!h]
  \centering
  \epsfig{file=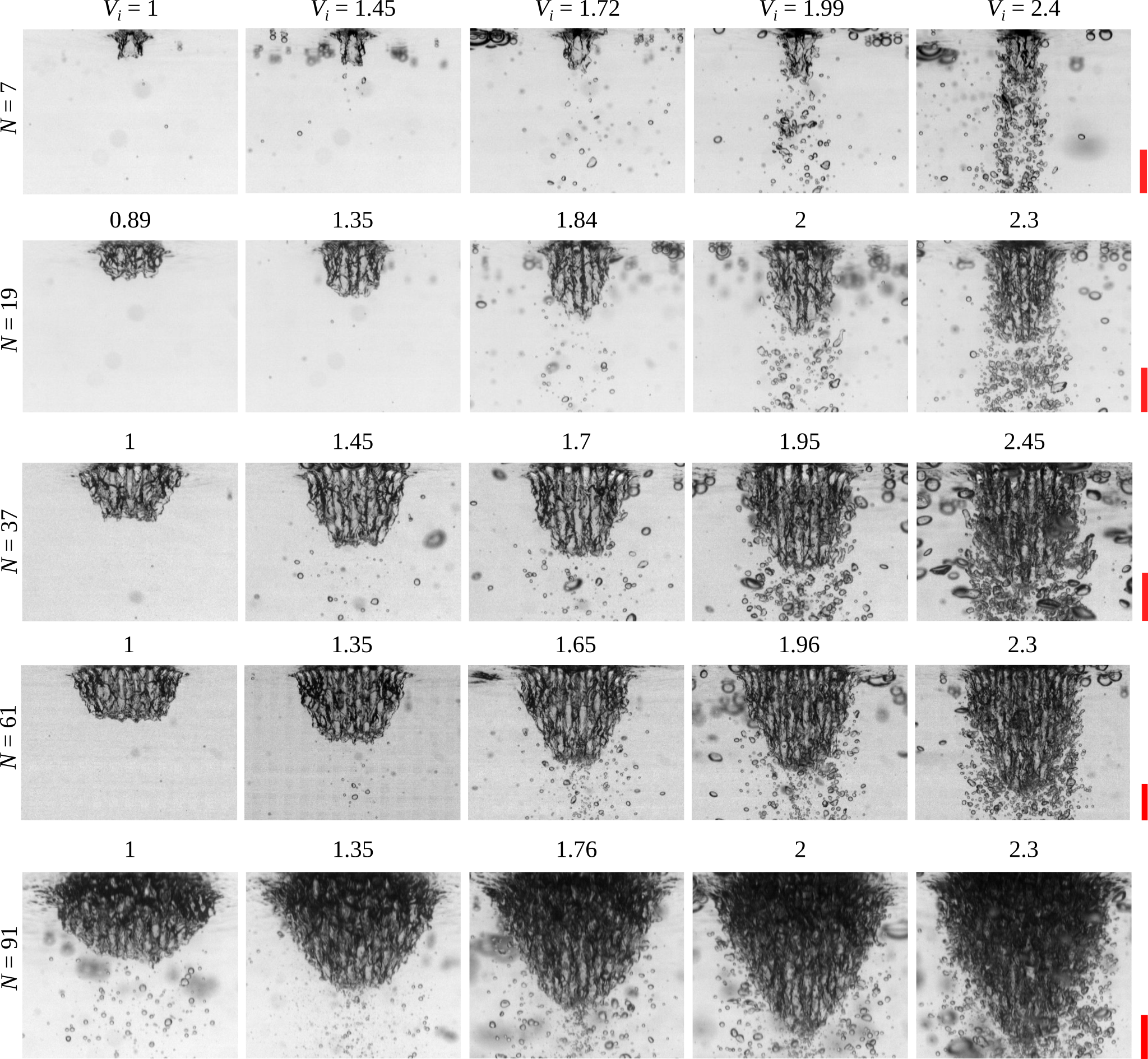,width=1\textwidth,keepaspectratio=true}
  \caption{{Images of air-cavity just below the free surface for multi-plunging jets as a function of number of jets ($N$) and impact velocity ($V_i$) for a fixed spacing $X/D_n = 1.5$. 
  The red scale bar indicates $1$~cm.}}
\label{fig:N_Dome}
\end{figure}

At a given speed, say $V_i \approx 1$~m/s, the height and width of the cavity increases as the number of jets $N$ is increased. \jon{Under identical jet conditions, say $V_i \approx 2$~m/s, the dome height $H_{Dome}$  in the case of $N =  91$ is approximately $10$ times greater than that observed for a single jet.} 
Figure \ref{fig:N_Dome} also evidences a significant role of impact velocity $V_i$ for fixed $N$. At the smallest velocity shown here for $N = 19$ jets, the dome resembles a flat pancake with distinct spikes. {At higher speeds $V_i$}, the cavity depth $H_{Dome}$ increases  and progressively sharpens at its centre. Correspondingly, bubble formation characteristics evolve significantly with increasing velocity. This is readily visible from the bubble sizes observed in all photographs of Figure \ref{fig:N_Dome}. {At the lowest impact velocities shown here, only very small bubbles form from thin fingers at the base and from thin air sheets.} Most of the fingers on the dome typically exhibit elongation and retraction without significant bubble detachment {(see Supplemental Video IV)}. With a moderate increase in jet speed, slightly larger bubbles also begin detaching from the base and occasionally from lateral fingers. At higher velocities, larger bubbles form more uniformly across the dome surface. At {the largest velocity ($2.3$~m/s) presented here}, bubble formation also initiates from the dome periphery, \naren{mostly due to the enhanced corrugation of each plunging jet}. It is pointed out here that the images displayed in Figure \ref{fig:N_Dome} were taken after the establishment of stationary flow, but before rising bubbles released at the foot of the cloud reached the dome. As $V_i$ increases, the dome is less discernible -- as for example in $N \leq 61$ at about $2.4$~m/s.  
This obstruction of the dome by bubbles is even more pronounced 
when the fall height is increased, leading to {a vigorous} bubble production, as in the case of single circular jet. 

\begin{figure}[!h]
\centering
  \epsfig{file=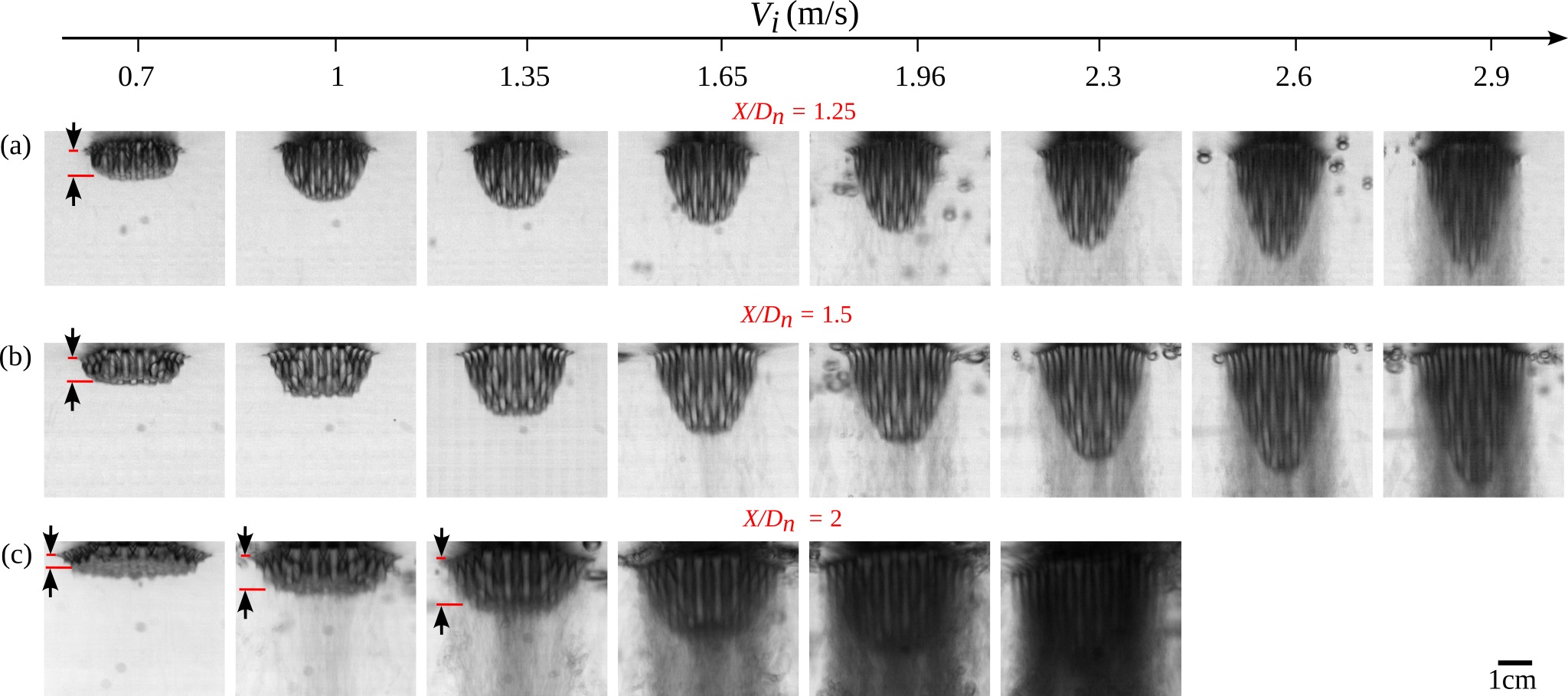,width=1\textwidth,keepaspectratio=true}
  
  \caption{\jon{Intensity-averaged dome image with impact velocity $V_i$ for different spacing values $X/D_n$ ($N = 61$). Arrows indicate the measured dome height, shown to avoid parallax effects.}}
\label{fig:projected_images}
\end{figure} 
\begin{figure}[!h]
\centering
  \epsfig{file=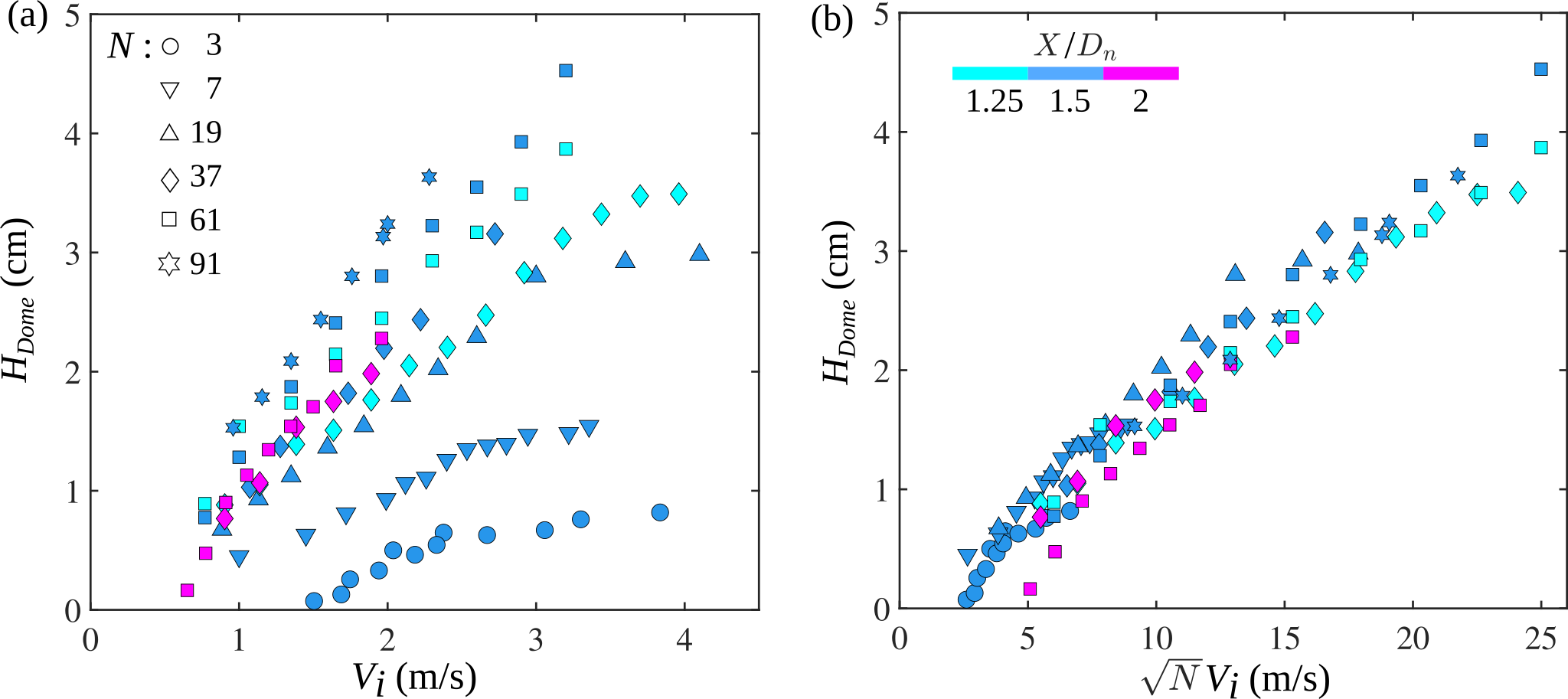,width=0.98\textwidth,keepaspectratio=true}
  \caption{{Evolution of dome height $H_{Dome}$ for various $N$ values and jet spacing $X/D_n$ as a function of (a) impact velocity $V_i$ and (b)  $\sqrt{N}V_i $, which is expected to control the individual jet expansion angle under the surface.}}
\label{fig:Height_graphs}
\end{figure}

\begin{figure}[!h]
    \centering
  \epsfig{file=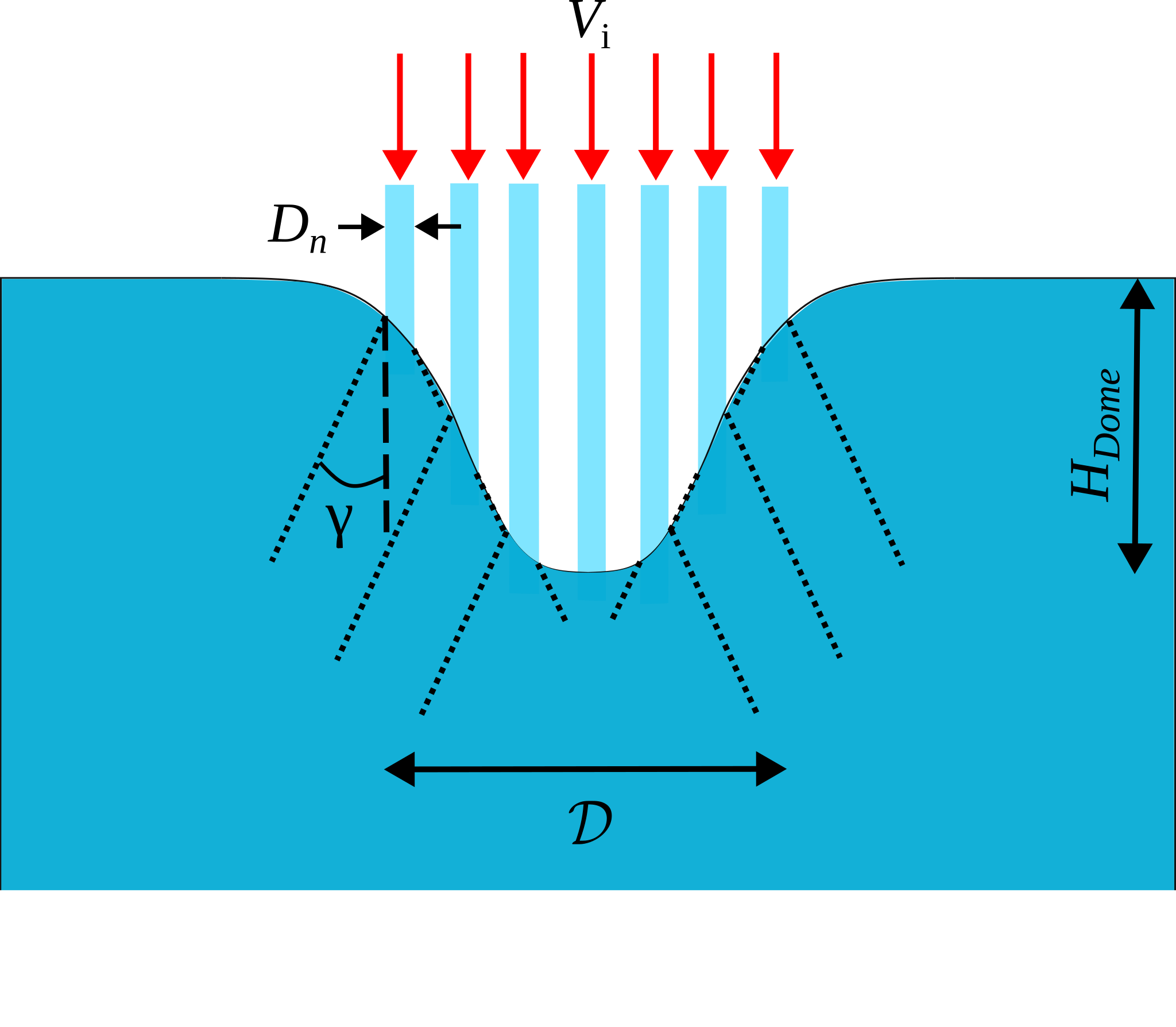,width=0.5\textwidth,keepaspectratio=true}
    \caption{Schematic of the  model for dome formation beneath an ensemble of impinging jets. Identical jets of diameter $D_n$ impact the free surface with velocity $V_i$, entraining liquid from the bath and inducing a successive radial expansion and deceleration of the submerged jets. The collective momentum redistribution and liquid entrainment drive the formation of a coherent dome of depth $H_{\mathrm{Dome}}$. 
 }
    \label{fig:model}
\end{figure}

\naren{Figure~\ref{fig:projected_images} presents intensity-averaged images of the dome structure, constructed from approximately $1000$ images. These time-averaged images closely resemble the dome shape observed through the naked eye. The structure is composed of multiple tubes with oval ends, corresponding to the individual jets viewed through the surrounding water, as discussed in Section~\ref{sec:cavity_exp}. The dome height is extracted from these averaged images. At higher impact velocities, intense bubble production significantly darkens the intensity-averaged images, making it difficult to distinguish the dome boundary reliably. Consequently, data are not extracted for these conditions. Figure~\ref{fig:Height_graphs}(a) illustrates the corresponding time-averaged dome height $H_{\mathrm{Dome}}$ as a function of the jet impact velocity $V_i$ for various configurations, namely, number of jets ($N$) and spacing between individual circular jets $X/D_n$ in the hexagonal arrangement.  The data consistently show that the average dome height increases with impact velocity. For fixed $V_i$ and $X/D_n$, $H_{\mathrm{D}ome}$ increases monotonically with the number of jets $N$.  Whereas for fixed $V_i$ and $N$ {(say $61$)}, the dependence of $H_{\mathrm{D}ome}$  on $X/D_n$ is non-monotonic, with $H_{\mathrm{D}ome}$ first increasing and then decreasing as $X/D_n$ increases. This behavior indicates that jet--jet coupling weakens beyond an optimal spacing. This trend is consistent with the observation that no common cavity was observed in \citet{dev2025} for the case of three jets at $X/D_n=5$. It should be noted, however, that this non-monotonic trend is observed only for impact velocities exceeding approximately 1 m/s.  At lower velocities, the $N$ = 61 triplets show a consistent decrease of $H_{\mathrm{D}ome}$
 with increasing $X/D_n$. This behaviour may be attributed to spacing-dependent changes in dome morphology. As shown by the intensity-averaged images for $N$ = 61 in Figure \ref{fig:projected_images}, the dome transitions at low velocities from a compact, \textit{pyramidal} structure at small $X/D_n$ to a shallow \textit{pancake-like} shape at larger $X/D_n$. At higher impact velocities, the dome geometry becomes similar across different $X/D_n$ values, and the non-monotonic dependence on $X/D_n$ is more clearly observed. A more detailed assessment of this transition would require a larger experimental dataset and complementary numerical simulations.}

For high-viscosity jets,  \cite{lorenceau2004air} proposed a scaling $L \sim \sqrt{\mu V_i/\rho g}$  for the extent down to which a cusp-like cavity dips into the pool before the threshold for air entrainment is reached, based on the competition between viscous drag and gravity. Here, $\mu$ and $\rho$ are the liquid dynamic viscosity and density, respectively. Such a scaling clearly under-predicts the dome height for the much lower viscosity conditions considered here. 

We suggest that the size of the cavity is actually controlled not by air entrainment, but by  \textit{liquid} entrainment, which occurs when the plunging jets entrain the liquid from the bath. When the $N$ jets hit the surface in steady flow, the outer ring  pulls in liquid from the bath, and the outer jets  therefore expand as they slow. \naren{At some depth this outer ring will reach the neighbouring smaller ring, which will itself start to slow and expand, and so on until the central jet is reached at the bottom of the dome, see Figure \ref{fig:model}. This suggests that the average slope of the dome $H_{Dome}/\mathcal{D}$ is directly controlled by the angle of expansion of each single jet $\gamma$. More precisely, $H_{Dome}/\mathcal{D} \approx 1/ \tan \gamma$. 
For each submerged jet, we expect $\tan \gamma$ to be inversely proportional to the square root of the jet momentum \citep{Schlichting2000}, and therefore to the jet velocity $V_i$.
This predicts $H_{Dome} \propto \mathcal{D}V_i \propto \sqrt{N}V_i$.} 
{When the cavity depth is plotted against this quantity as in Figure~\ref{fig:Height_graphs}(b), measurements collapse about a common trend. Despite some scatter, this is striking given the large variation of $H_{Dome}$ when experimental parameters were varied.} \naren{We believe this mechanism is also consistent with the impact of jet spacing $X$ on the shape of the dome, as seen in Figure \ref{fig:projected_images}. For smaller $X$, the inner jets are well shielded from the bath by the outer jets, and the tip of the dome is visible, corresponding to the depth at which the central jet starts entraining liquid. When $X$ is increased, though, the shielding effect of the core jets is weakened, liquid entrainment occurs more easily, resulting in the observed pancake shape.}

A more rigorous expression for $H_{Dome}$ should include a non-zero mean void fraction $\phi_0$ at impact. This might also contribute to the scatter for different values of $X/D_n$, and also for higher $V_i$ at fixed $X/D_n$. 

\jon{The inverted dome-shaped air-water cavity is not expected to form, if the jet spacing $X/D_n$ become very large. In our previous study on triple jet interactions \citep{dev2025}, we observed that at $X/D_n = 5$ the jets do not interact near the impact zone at all. Instead of a dome, individual conical-shaped bubble clouds interact further downstream in the bubble cloud region \citep[see Fig. 3]{dev2025}. Figure~\ref{fig:xDome} presents visualisations taken from injector plates corresponding to $X/D_n = 3$ and $X/D_n = 5$. When both cases are compared to Figure~\ref{fig:dome61}, a dense bubble cloud closer to the impact zone is observed. Large bubbles circulate inside the cloud before escaping it (Supplemental Video VII). However, a closer look at the air entrainment just below the jet impact shows bubbles that emanate from very small cavities around individual jets, as if each jet entrains air independent of each other. The dome has not shrunk, but it simply does not occur in these cases.}
\begin{figure}[!h]
\begin{center}
  \epsfig{file=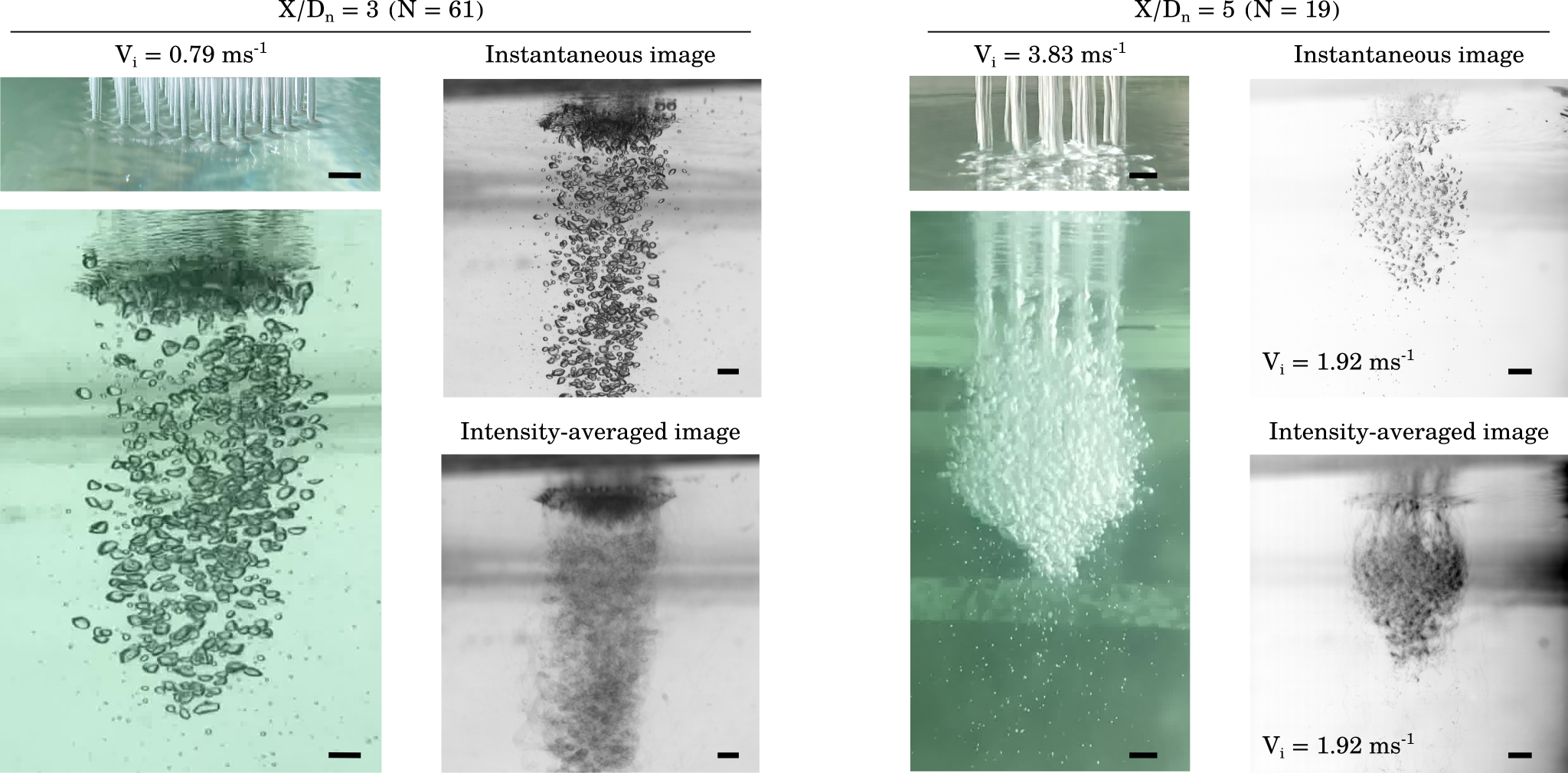, width=1\textwidth,keepaspectratio=true}
\caption{\jon{Effect of $X/D_n$ on the dome formation at $Z_f = 2$~cm (see Supplemental Video VII). Color and BW images were acquired at $240$ FPS and $3000$ FPS, respectively. Scales indicate $1$~cm.}}
\label{fig:xDome}
\end{center}
\end{figure}

\jon{In general, a dome forms a sealed cavity at the air-water interface, whose walls and seals are respectively provided by the impinging jets and the interface spanning the gaps between neighbouring jets. The interface can retain air only as long as its curvature counteracts the hydrostatic pressure of the surrounding bath. This balance is determined by the capillary length $l_c = 2.7$~mm. For jet-to-jet gaps, say $w = X - D_n$, narrower than this value, the interface acts as a rigid cap that sustains a pressure difference of approximately $2\sigma/w$, effectively sealing the cavity. For wider gaps, the interface flattens, and the pressure difference to sustain the air-water cavity decreases. Subsequently, air escapes into the liquid bulk at the rate at which the jets entrain it.  Therefore, the criterion $w \lesssim l_c$, or equivalently $X/D_n \lesssim 1 + l_c/D_n \approx 2$, determines the dome formation for the nozzles studied here. Our data support this argument, as domes form at gaps when $w/l_c \simeq 0.25$, $0.50$, and $1$ (at $X/D_n =$~$1.25$, $1.5$, $2$, respectively). In contrast, domes do not form at $w/l_c \simeq 2$ and $4$ (or correspondingly, $X/D_n = 3$, $5$, respectively). The marginal case at $X/D_n = 2$, where $w \approx l_c$, exhibits the flattest and smallest domes.}

\jon{At least three geometric length scales are relevant among the control parameters, namely, the jet diameter $D_n$, the inter-jet spacing $X$, and the fall height $Z_f$. The inter-ring distance could be modified as well. But evaluating each parameter is beyond the present scope. Figure~\ref{fig:Height_graphs} demonstrates that \textit{cone-shaped} liquid entrainment alone can capture the first-order effect of the jet spacing $X/D_n$ for dome's height within the regime where a dome forms. The cases $X/D_n = 3$ and $X/D_n = 5$ in  Figure~\ref{fig:xDome} strongly suggest that the dome disappears when the distance between neighbouring jets is larger than the capillary length.}

\section{Conclusion}
We considered model experiments to investigate the impact site formed by a set of closely-packed water jets plunging into a pool of water. 
Our experiments show that the impact site due to multiple plunging jets is distinctly different from the cavity around either a single jet impact or a drop-train impact. We observed the presence of a novel inverted dome-shaped air-water cavity formed at the impact site for all multi-jet setups. We used phase-detection optical probes to measure local air-to-water fraction and deduced that the cavity contains multiple stems and pockets of air, while successive rings of water jets mix with the pool water at different depths. We also presented various bubble production mechanisms at different locations inside the cavity. 
{Bubbles of sub-millimetre and millimetre sizes are formed at the air cavity in the dome periphery. Between two rings of water jets, thin sheets of air are observed to stretch, rapidly thin-out and break-up into $100$~$\mu$m size bubbles.}
Finally, we proposed a scaling law with which it is possible to collapse the average dome heights observed when the number of jets and the impact velocity are varied in a wide range. \naren{The proposed mechanism for the observed shape of the dome is based on liquid entrainment in the otherwise quiescent pool by concentric rings of moving liquid jets. The outer jets initially shield the inner jets, and liquid entrainment at the center of the arrangement can only occur once outer jets have expanded, until a single larger liquid jet has been formed below the surface.} By increasing the inter-jet spacing $X$, we showed that the dome disappears when $X/D_n > 2$.
When very large water bodies freely fall in natural or industrial situations, they break into multiple jets, which hit the liquid surface randomly. Even though the dome structure and its associated bubble production mechanisms have been described here in a stationary regime, it is very likely that they also exist in any transient configuration involving liquid entrainment between neighbouring jets. 

{\textbf{Acknowledgements.} We thank N. Grosjean for LIF imaging rig, and A. Buridon, G. Geniquet and S. Martinez for technical assistance. We also thank A2 Photonic Sensors for the support with the optical probes. 

\textbf{Funding.} This research was funded by the French Agence Nationale de la Recherche ANR under grant no. JETPLUME ANR-21-CE05-0029-01.}

\textbf{Declaration of interests.} The authors report no conflict of interest.

\bibliographystyle{plainnat}
\bibliography{BiblioMultiJetDome}

\end{document}